\documentclass{ACAT}
\usepackage{graphicx, wrapfig}
\usepackage{caption, subcaption}
\usepackage{booktabs}
\usepackage[numbers]{natbib}
\begin{document}

\title{HEP Benchmark Suite: Enhancing Efficiency and Sustainability in Worldwide LHC Computing Infrastructures}

\author{Natalia Szczepanek$^{1}$, David Britton$^{2}$, Alessandro Di Girolamo$^{1}$, Ewoud Ketele$^{1}$, 
Ivan Glushkov$^{3}$, Domenico Giordano$^{1}$, Ladislav Ondris$^{1}$, Emanuele Simili$^{2}$, Gonzalo Menendez Borge$^{1}$}

\affil{$^1$CERN, European Laboratory for Particle Physics, Geneva, Switzerland}
\affil{$^2$University of Glasgow, Glasgow, United Kingdom}
\affil{$^3$University of Texas, Arlington, United States}

\email{natalia.diana.szczepanek@cern.ch, domenico.giordano@cern.ch}

\begin{abstract}
As the scientific community continues to push the boundaries of computing capabilities, there is a growing responsibility to address the associated energy consumption and carbon footprint. This responsibility extends to the Worldwide LHC Computing Grid (WLCG), encompassing over 170 sites in 40 countries, supporting vital computing, disk, tape storage and network for LHC experiments. Ensuring efficient operational practices across these diverse sites is crucial beyond mere performance metrics.
This paper introduces the HEP Benchmark suite, an enhanced suite designed to measure computing resource performance uniformly across all WLCG sites, using HEPScore23 as performance unit. The suite expands beyond assessing only the execution speed via HEPScore23. In fact the suite incorporates metrics such as machine load, memory usage, memory swap, and notably, power consumption. Its adaptability and user-friendly interface enable comprehensive acquisition of system-related data alongside benchmarking.
Throughout 2023, this tool underwent rigorous testing across numerous WLCG sites. The focus was on studying compute job slot performance and correlating these with fabric metrics. Initial analysis unveiled the tool's efficacy in establishing a standardized model for compute resource utilization while pinpointing anomalies, often stemming from site misconfigurations.
This paper aims to elucidate the tool's functionality and present the results obtained from extensive testing. By disseminating this information, the objective is to raise awareness within the community about this probing model, fostering broader adoption and encouraging responsible computing practices that prioritize both performance and environmental impact mitigation.
\end{abstract}

\section{Introduction}
The Worldwide LHC Computing Grid (WLCG) contain more than 170 sites in over 40 countries providing computing, disk, tape storage and network to the LHC experiments. The WLCG adopted HEPScore23 (HS23) \cite{chep:hs23} as a performance metric for resource capacity planning, hardware acquisition, pledging of future resources and usage accounting of the experiments. HEPScore is a tool that coordinates the consecutive execution of various containerised HEP-Workloads containers and determine the final benchmark score of a given compute resource \cite{springer:2021:article}. HEPScore is orchestrated by the HEP Benchmark Suite, which also enables the execution for non-HEP benchmarks such as HEP-SPEC06 (HS06) \cite{physics:hs06}, SPEC CPU 2017 \cite{speccpu2017}, and DB12 \cite{Charpentier2017}. HS23 includes seven workloads from five experiments, each running with 3 repetitions with the latest version of the experimental software, which supports both x86\_64 and aarch64 architectures. This is particularly important as several new architectures are expected to emerge in the future alongside x86\_64 currently used in WLCG. The reference machine for this benchmark is an Intel® Xeon® Gold 6326 CPU @ 2.90 GHz with hyperthreading (HT) enabled \cite{springer:2021:article}. The Suite can be easily installed and used as described in \cite{hepix:benchmarking-how-to}.
\begin{figure}[ht]
    \centering
    \includegraphics[width=0.8\linewidth]{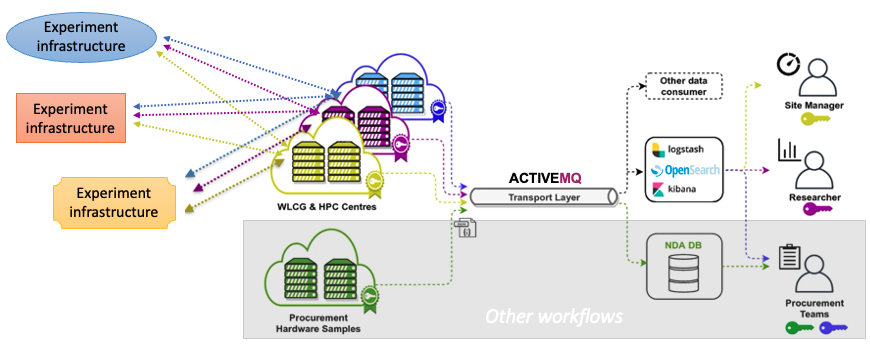}
    \captionsetup{width=1\linewidth}
    \caption{The HEPiX Benchmarking Solution for WLCG Computing Resources provides a reusable infrastructure suitable for various workload management systems. The HEP Benchmark Suite integrates with any job submission infrastructure, allowing jobs to be submitted to selected sites. Once the job containing the benchmarking script is executed, the results are automatically sent via message broker to dedicated database. From there, site managers, researchers, and other users can retrieve and analyze the results.}
    \label{fig:infra}
\end{figure}
Thanks to the design of the Suite, it is possible to publish the results to dedicated database; and then to perform data analysis using visualization tools such as Grafana \cite{grafana}. HEPScore can be adjusted to experiments needs and be used not only on bare-metal nodes \cite{wiki:baremetal} and local machines, but also in automated or semi-automated way on WLCG batch system, by combining the Suite with the experiments job submission infrastructures, which is conceptually presented in \textbf{Figure~\ref{fig:infra}}. It was positively tested with the ATLAS \cite{experiment:atlas} and LHCb \cite{experiment:lhcb} collaborations. This approach can scale up the data acquisition of benchmarking data and allow more accurate and broad analysis.

Continuous acquisition and monitoring of benchmark data offer significant benefits for experiments. This process aids in monitoring machine performance, facilitates decision-making regarding site utilization, and assists in resolving configuration issues. Site administrators can leverage this system to oversee their infrastructure, while experiments can utilize the data to optimize site usage and detect any misconfigurations. The HEPiX Benchmarking Working group extended use of the HEP Benchmark Suite beyond conventional boundaries. Despite the existing capability to capture performance and diverse system information, the group designed a highly adaptable and user-friendly methods for acquiring various system-related metrics alongside the benchmark. The Suite Plugins can be easily extended and tailored by new metrics depending on user needs, thanks to its flexible and intuitive modification. Collected metrics can be compared with the data from reference machines, enabling the detection of anomalies associated with system misconfiguration.

\section{HEP Benchmark Suite with Plugins}

\begin{figure}[t]
    \centering
    \begin{subfigure}[t]{0.52\textwidth}
        \centering
        \includegraphics[width=\linewidth]{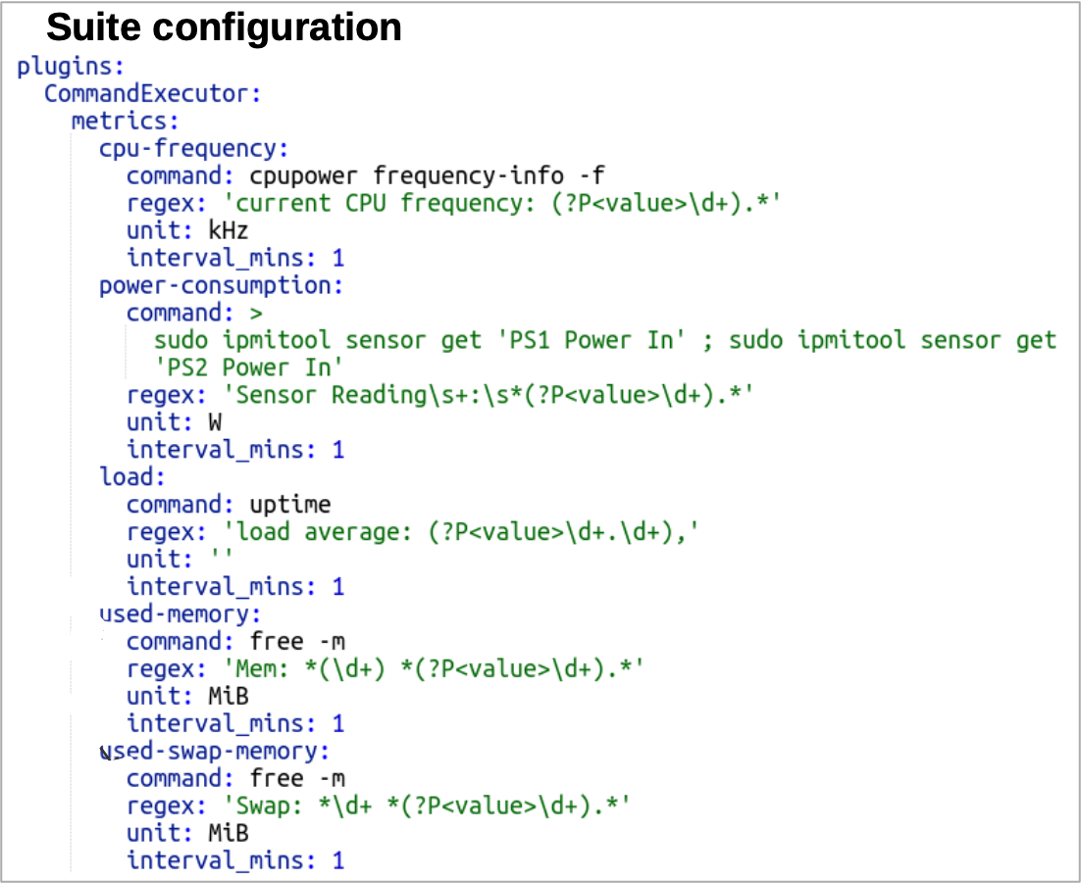}
        \caption{The Command Executor Plugins facilitate metric definition by allowing users to specify the metric name, command, regex, unit, and interval (in minutes, interval\_mins).}
        \label{fig:suite-configuration}
    \end{subfigure}
    \hfill
    \begin{subfigure}[t]{0.46\textwidth}
        \centering
        \includegraphics[width=\linewidth]{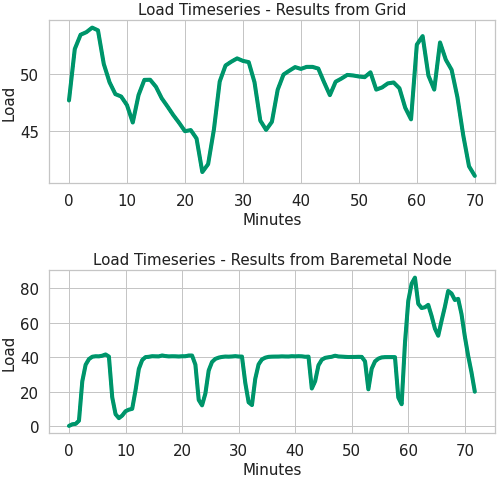}
        \caption{The timeseries data from the JSON file can be visualized to analyze performance. The top section shows load from a grid environment, while the bottom section illustrates data from a bare-metal node, highlighting differences between isolated test results and those observed in a production environment.}
        \label{fig:plugins-timeseris}
    \end{subfigure}
    \caption{HEP Benchmark Suite Plugins: Configuration and timeseries data collection. (a) Configuration of the HEP Benchmark Suite Plugins is easy to modify by the user. Data collection can be performed on bare-metal nodes (lower (b)) as well as on grid (upper (b)). When executing on a grid, the collected data reflects the overall load on the entire machine, rather than just the individual job slot.}
    \label{fig:plugins}
\end{figure}

The Suite Plugins were developed to introduce flexible and extensible functionality to the Suite, independent of the benchmark, and to gather additional data such as machine load, power consumption, and memory usage. The design allows for independent operation alongside the benchmark, ensuring the execution time remains unchanged, and enables data collection in three phases: before (pre), during, and after (post). The “pre” and “post” phases run only for a brief configurable moment to retrieve a single measurement, which can be modified by the user. The ease of use and extended functionality of the HEP Benchmark Suite are attributed to the Command Executor Plugin, which can be seen in \textbf{Figure~\ref{fig:suite-configuration}}. The modification and addition of new metrics are highly flexible. Each metric has several parameters, such as command to be executed, regular expression (regex) to extract the relevant metric, unit, interval\_mins, and aggregation, that can be adjusted based on user needs. In \textbf{Figure~\ref{fig:plugins-timeseris}}, the timeseries of the load, as one of the collected metric is visualized.

\section{Data Analysis}
A significant amount of data has been collected using the HEP Benchmark Suite with Plugins for different workload management systems (WMS) such as those of ATLAS and LHCb. The statistics of the collected data are presented in \textbf{Table~\ref{tab:statistics}}. For LHCb, because most of its job slots are single core and cannot run the multicore WLs of HS23, a dedicated workload was submitted with three repetition on a single core job slot, manually, using DIRAC \cite{lhcb:dirac}. For ATLAS, extensive data was gathered using the automated job submission capabilities of HammerCloud \cite{atlas:hammercloud} and PanDA \cite{atlas:bigpanda}. The official HS23 configuration was utilized, with each job slot running on 8 cores and a single repetition. This strategy was adopted due to its shorter execution times and automated processes, enabling the collection of a substantial volume of statistical data. This study focuses on analyzing the machine's load and its relationship to performance, highlighting the most noteworthy and significant aspects of the gathered data.
\begin{table}[hb]
  \centering
  \caption{Data Statistics for each of the experiments. Automated submission for ATLAS allows the collection of extensive statistics data every day. The walltime shown for ATLAS is the effect of running one repetition of the full HS23 benchmark. For LHCb, just the dedicated LHCb benchmark workload was run, three times.}
  \begin{tabular}{c c c c c c}
    \toprule
    \textbf{Experiment} & \textbf{No. \#} & \textbf{No. Sites} & \textbf{No. CPU Models} & \textbf{Walltime [min]} & \textbf{\% of total walltime} \\
    \midrule
    ATLAS & 148257 & 139 & 227 & 81 & 0.06\\
    \midrule
    LHCb & 2100 & 48 & 110 & 43 & N/A \\
    \bottomrule
  \end{tabular}
  \label{tab:statistics}
\end{table}
\subsection{HS23 vs Load analysis}
\label{subsec:load-measurements}
\begin{figure}[h]
    \centering
    \begin{subfigure}[t]{0.52\textwidth}
        \centering
        \includegraphics[width=\linewidth]{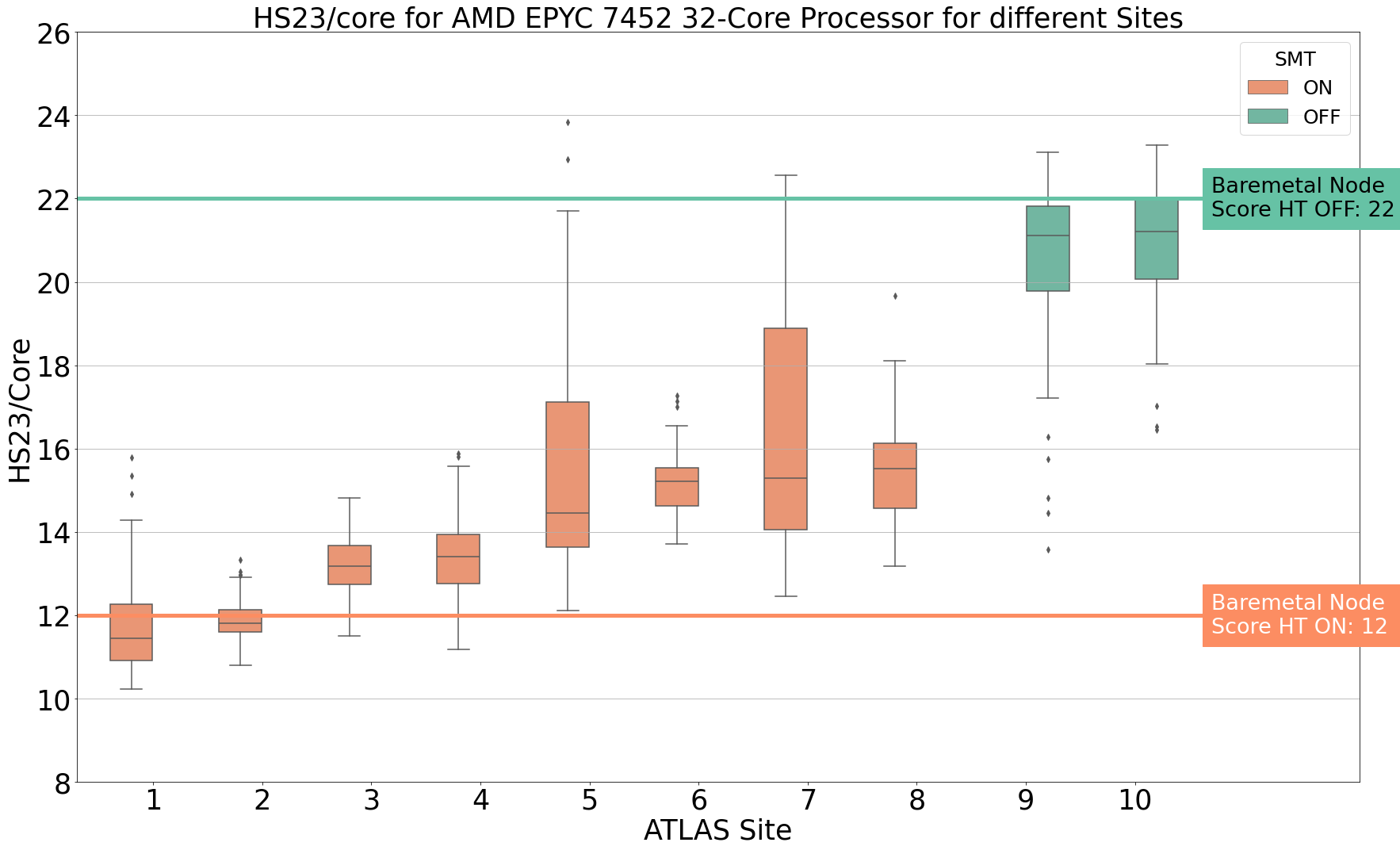}
        \caption{A boxplot displays HS23 per physical core data across various (ATLAS) sites (numbered 1 to 10), distinguishing between HT ON and OFF conditions. The results range from approximately 11 to 19 for HT ON, with the lines representing data collected from bare-metal nodes during the standard benchmarking procedure.}
        \label{fig:grid-results}
    \end{subfigure}
    \hfill
    \begin{subfigure}[t]{0.46\textwidth}
        \centering
        \includegraphics[width=\linewidth]{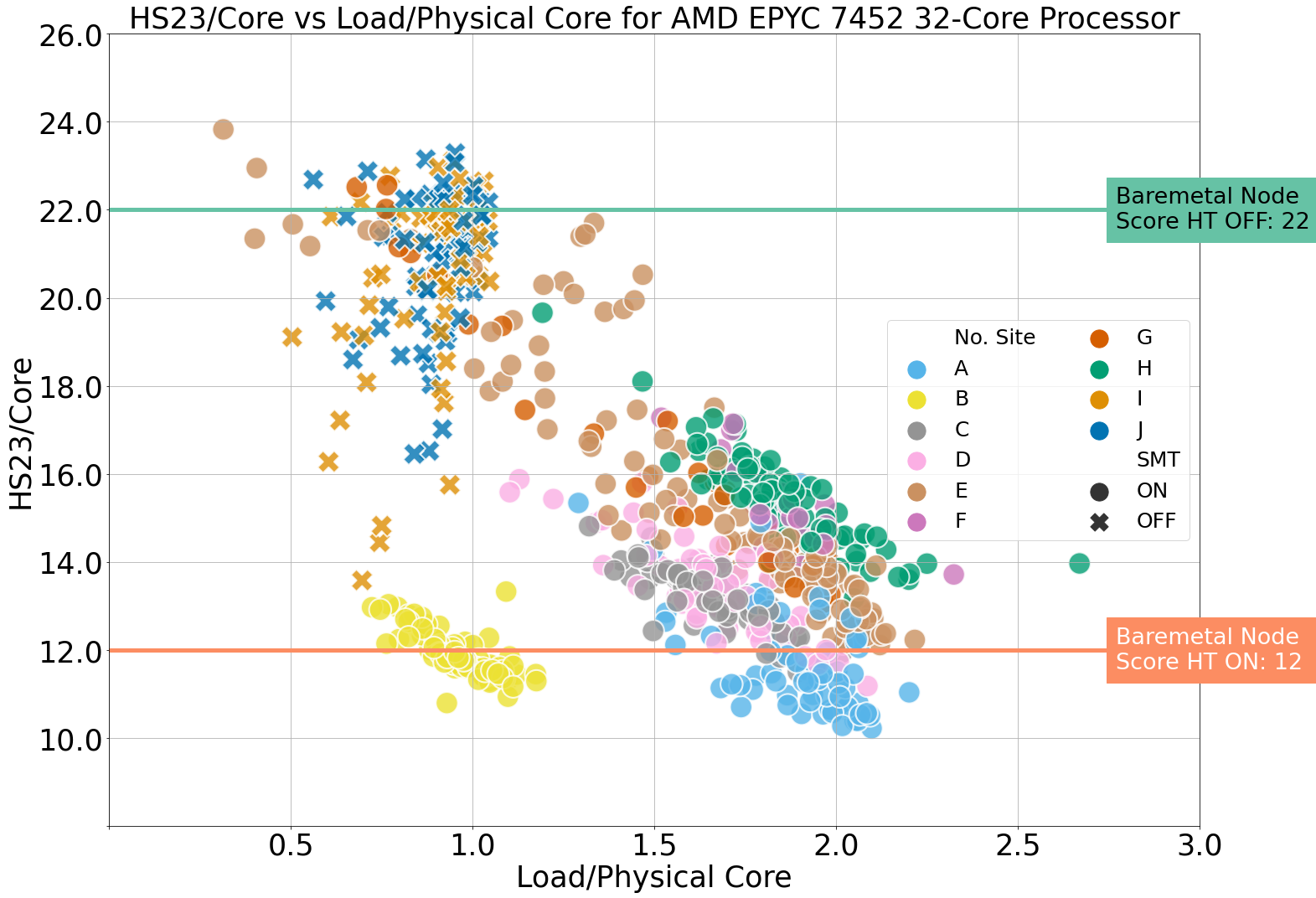}
        \caption{The scatter plot of HS23/Core versus Load/Physical Core illustrates how the performance of each site varies depending on the machine load at the time of benchmarking script execution.}
        \label{fig:grid-load-results}
    \end{subfigure}
    \caption{Results collected from different WLCG sites for AMD 7452 32-Core Processor CPU.}
    \label{fig:plugins-results}
\end{figure}

The idea of adding load to the collected metrics originated from analyzing results from the WLCG batch systems at multiple sites. The utilization of physical cores is imperative due to the availability of outcomes for both HT ON and OFF. In the context of logical cores, results would converge at a comparable level, introducing potential misinterpretation. It has been discovered that for the same CPU model, different sites achieved varying results, as illustrated in \textbf{Figure~\ref{fig:grid-results}}. These results were compared with those from bare-metal nodes, with the expectation that they would fall within a similar range. However, as presented, the differences between sites ranged from a few percent to almost 33\% when HT was enabled. Additionally, results within a single site, such as site 7, showed significant variability. The hypothesis was that these discrepancies were related to the load on the machines during the benchmarking script execution. 

After implementing the plugins functionality, the collection of the load of the machine has been possible, adding another dimension to our data. The updated data representation is shown in \textbf{Figure~\ref{fig:grid-load-results}}. As observed, performance variability is small for some sites, but significantly higher for others, showing correlation with the machine load. The general trend indicates that as the machine load increases, the HS23 per physical core decreases, supporting our hypothesis. Understanding the impact of machine load on performance variability is crucial for optimizing resource allocation and benchmarking accuracy in distributed computing environments. For instance, the significant variability observed at site 7 underlines the need for standardized load monitoring and management practices across sites to ensure consistent performance.

\begin{figure}[ht]
    \centering
    \begin{subfigure}[t]{0.49\textwidth}
        \includegraphics[width=\linewidth]{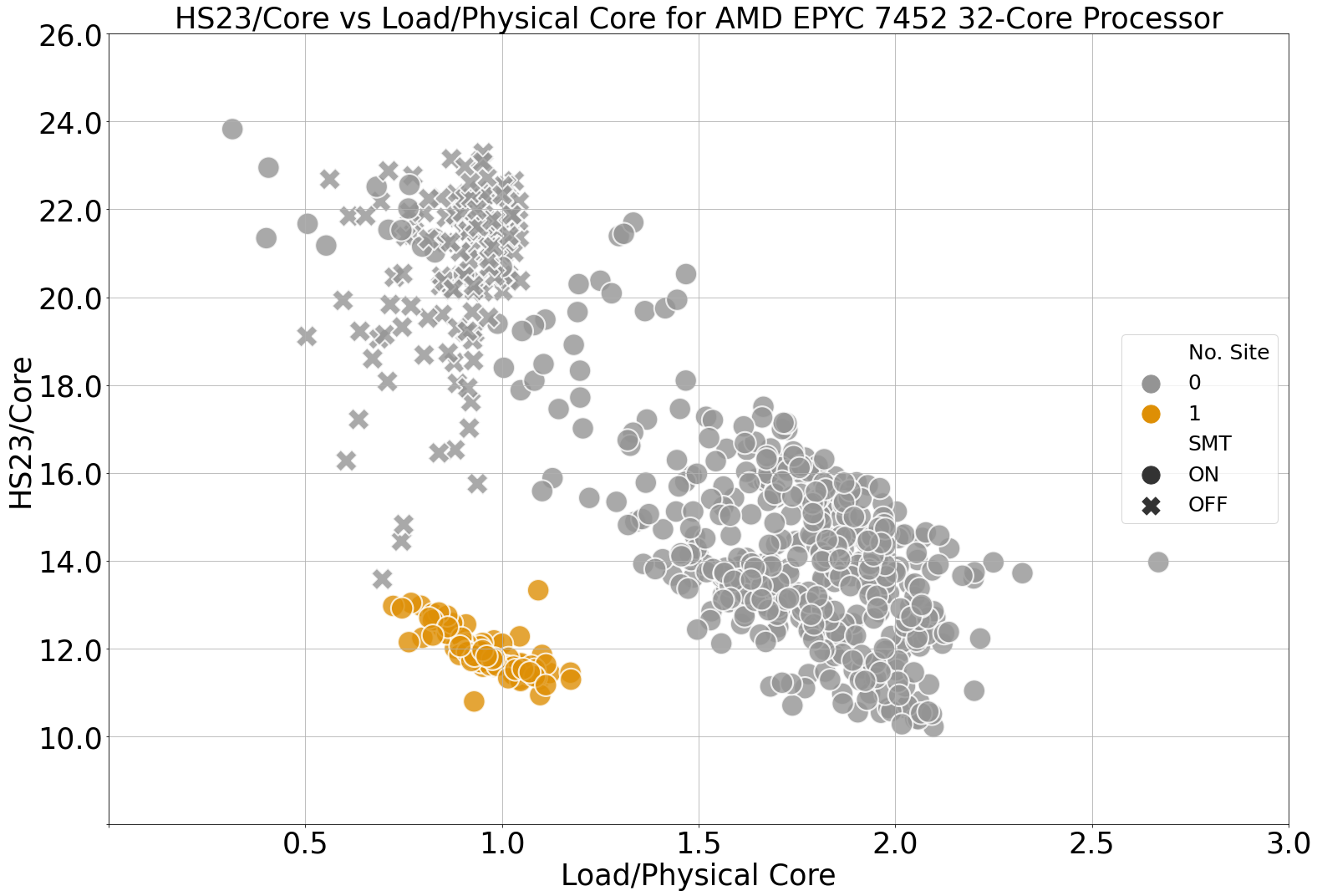}
        \caption{Before the intervention, the performance at the highlighted site, in terms of score per core, was only about half of that of other sites with the same load.}
        \label{fig:affected-before}
    \end{subfigure}
    \hfill
    \begin{subfigure}[t]{0.49\textwidth}
        \includegraphics[width=\linewidth]{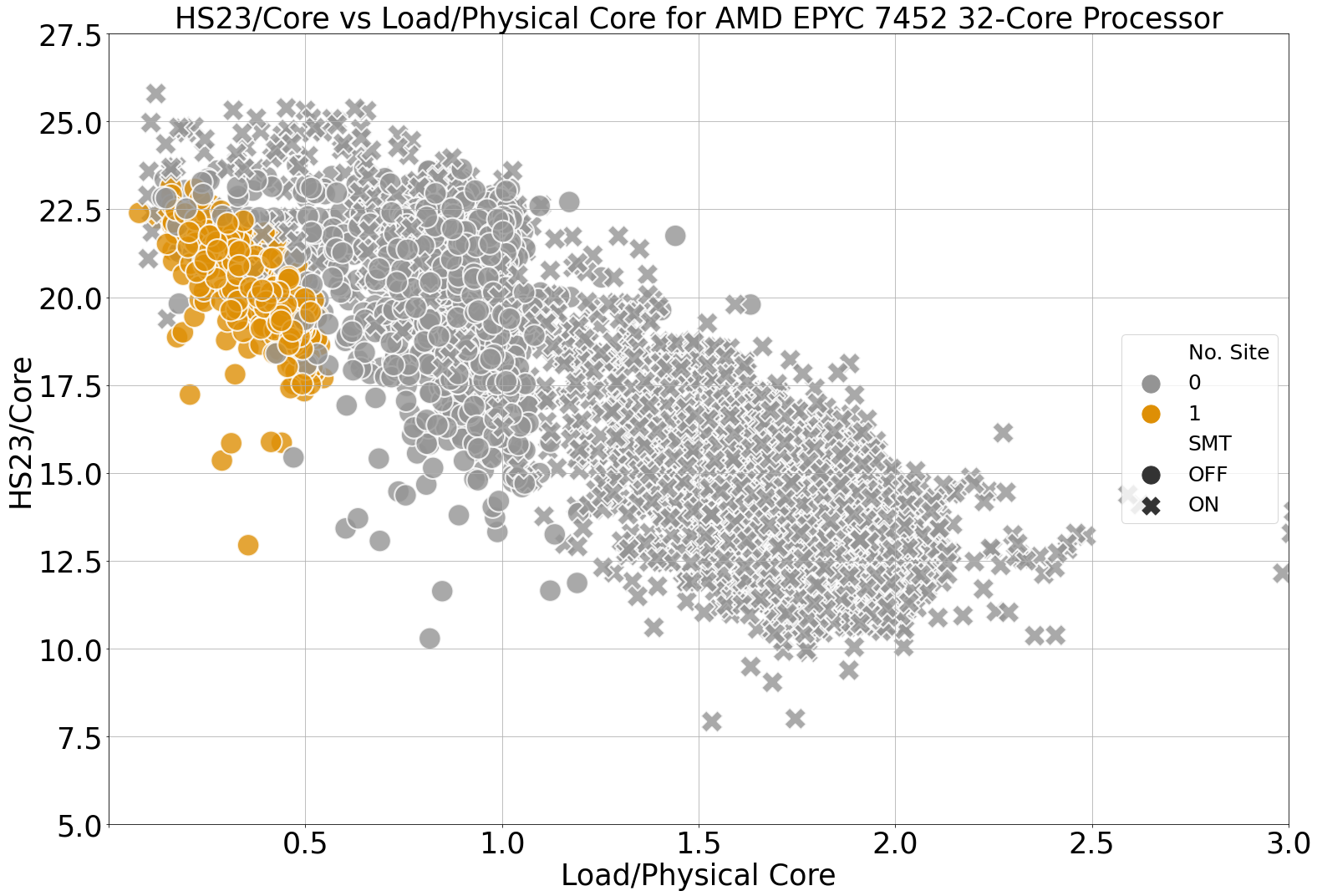}
        \caption{After the intervention, a 66\% improvement is observed at the highlighted site, demonstrating the effectiveness of the changes and the importance of proper configuration and optimization.}
        \label{fig:affected-after}
    \end{subfigure}
    \caption{Anomaly detection for AMD EPYC 7452 32-Core CPU reveals that the highlighted site exhibits performance below the expected trend (a). After the intervention performance increased by 66\% (b).}
    \label{fig:amd-measurements}
\end{figure}

The example of the AMD EPYC 7452 32-Core CPU is particularly intriguing when examining the overall trend, which indicates that as load increases, performance per core decreases. However, as shown in \textbf{Figure~\ref{fig:affected-before}}, there is a notable anomaly represented by the orange region, which diverges from this trend. Performance at this site, in terms of score per core, is notably lower compared to other sites with double the server load. Upon contacting the site administrator, it was discovered that the servers had incorrect configuration significantly impacting performance. Fixing the issue results in a 66\% improvement in performance, as depicted in \textbf{Figure~\ref{fig:affected-after}}. 

This outcome underscores how this analysis and corrective actions have enhanced site efficiency. Recursive studies across various CPU models identified and resolved additional configuration issues at different sites. These efforts also provided insights into the deliberate configuration choices made by site administrators, which, although initially appearing as anomalies, were intentional decisions. 

\subsection{Data Model and governor analysis}
\begin{figure}[ht]
    \centering
    \begin{subfigure}[t]{0.48\textwidth}
    \includegraphics[width=\linewidth]{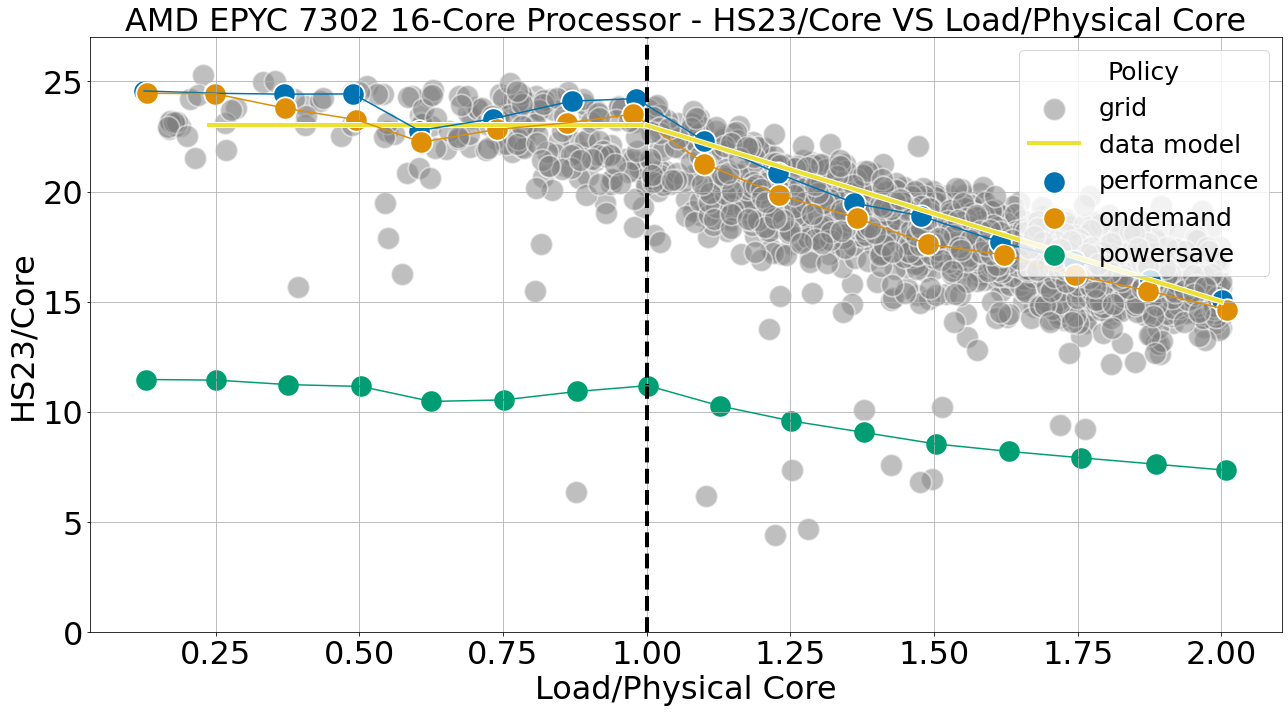}
    \caption{Pattern 1: Governor study on bare-metal node vs results from the grid vs data model for AMD EPYC 7302 16-Core Processor.}
    \label{fig:pattern1}
    \end{subfigure}
    \hfill
    \begin{subfigure}[t]{0.5\textwidth}
    \includegraphics[width=\linewidth]{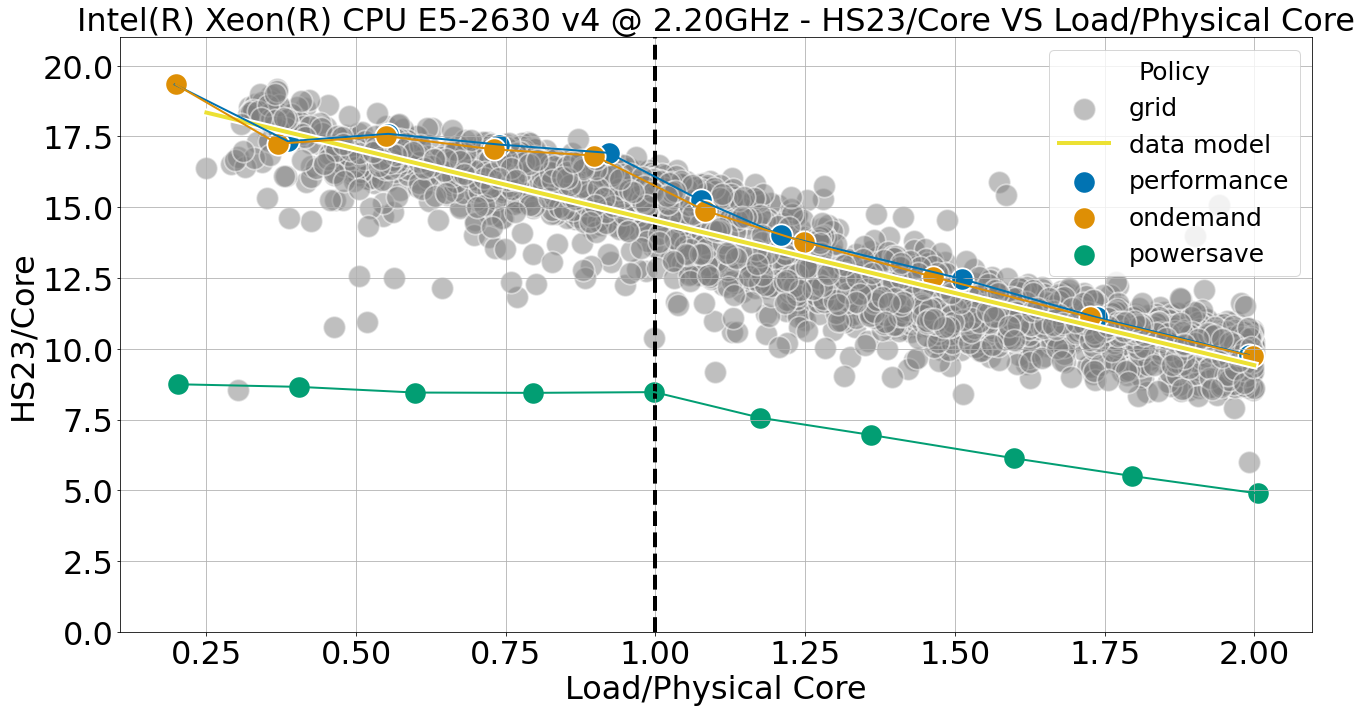}
    \caption{Pattern 2: Governor study on bare-metal node vs results from the grid vs data model for Intel® Xeon® CPU E5-2630 v4 @ 2.20GHz.}
    \label{fig:pattern2}
    \end{subfigure}
    \caption{Data from the grid shows two performance patterns in HS23 vs Load analysis. The first, observed in the low-load region (0,1), aligns with expected thread-scan behavior, as seen in (a) with stable performance under low load. The second pattern deviates from expectations, showing variable performance with load changes in (b). When using the 'powersave' governor, only the first pattern appears in isolated tests (a,b). In contrast, 'performance' and 'ondemand' governors exhibit both patterns consistently in isolated on bare-metal nodes and grid tests (a,b).}
    \label{fig:data-model}
\end{figure}

Thread scanning refers to the process of analyzing the performance of CPU threads under varying loads to identify patterns and anomalies. Based on the results of thread scan measurements and other related studies \cite{springer:2021:article}, a single pattern was expected in the data from the grid. However, two main patterns emerged: one where the load in the range (0,1) results in a constant performance despite increasing load (consistent with the thread scan) shown in \textbf{Figure~\ref{fig:pattern1}}, and another where performance decreases in this load range, as seen in \textbf{Figure~\ref{fig:pattern2}}. The presence of these two distinct patterns indicates the influence of hidden factors affecting performance. 

\begin{figure}[ht]
    \centering
    \includegraphics[width=0.95\linewidth]{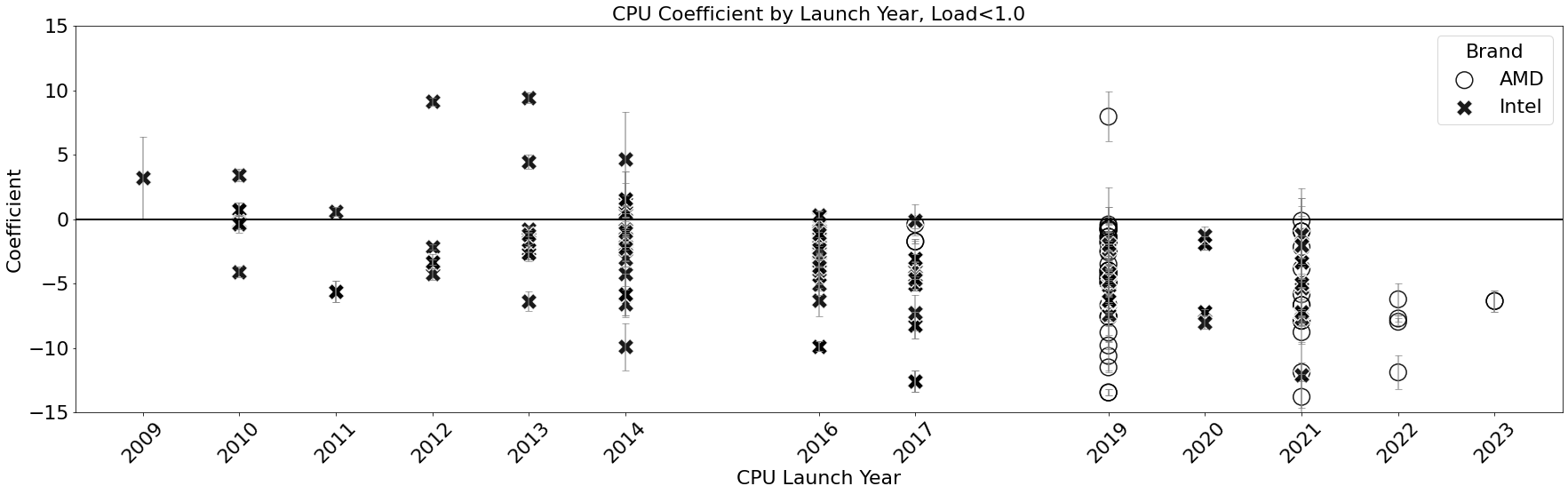}
    \caption{Coefficients derived from the fit line in the HS23 versus Load 2D plot, focusing on the low load region (0,1). Each point represents a different CPU model, categorized by brand and launch year.}
    \label{fig:coefficients}
\end{figure}

The initial goal was to determine whether the observed behavior was unique to the grid or also present on bare-metal nodes. Additionally, the impact of various governors \cite{redhat:governor} on system performance was investigated. Dedicated tests were conducted on bare-metal nodes using different governors, which are referred to later as isolated tests. The benchmarking was performed on AMD EPYC 7302 16-Core Processor and Intel® Xeon® CPU E5-2630 v4 @ 2.20GHz machines, employing a thread scan with a modularity of 4. The governors tested included performance, ondemand, and powersave. In \textbf{Figure~\ref{fig:pattern1}}, consistent performance trends were observed across all governors on bare-metal nodes, consistent with the first pattern observed on the grid. \textbf{Figure~\ref{fig:pattern2}} similarly illustrates performance characteristics on bare-metal nodes, showing decreased performance with increasing load, which is in line with the second visible pattern for the performance and ondemand governors. On the contrary, the powersave governor exhibited results aligning with the first pattern observed on bare-metal nodes, consistent with thread scan. Current hypothesis is that specific configuration aspects, such as turbo boost and thermal shielding effects associated with brand and CPU launch year, as illustrated in \textbf{Figure~\ref{fig:coefficients}}, may influence performance and contribute to the emergence of the second pattern. In this figure, coefficients are derived from the fit line in the HS23 versus Load within the low load region (0,1), where the patterns are observed. Each point represents a different (CPU Model, Site) pair, categorized by brand and launch year. The data suggest that newer architectures exhibit lower coefficients, which correlates with the appearance of the second pattern.
\subsection{Power Consumption and CPU Frequency Study}
\label{subsec:power-study}
\begin{figure}[ht]
    \centering
    \includegraphics[width=0.7\linewidth]{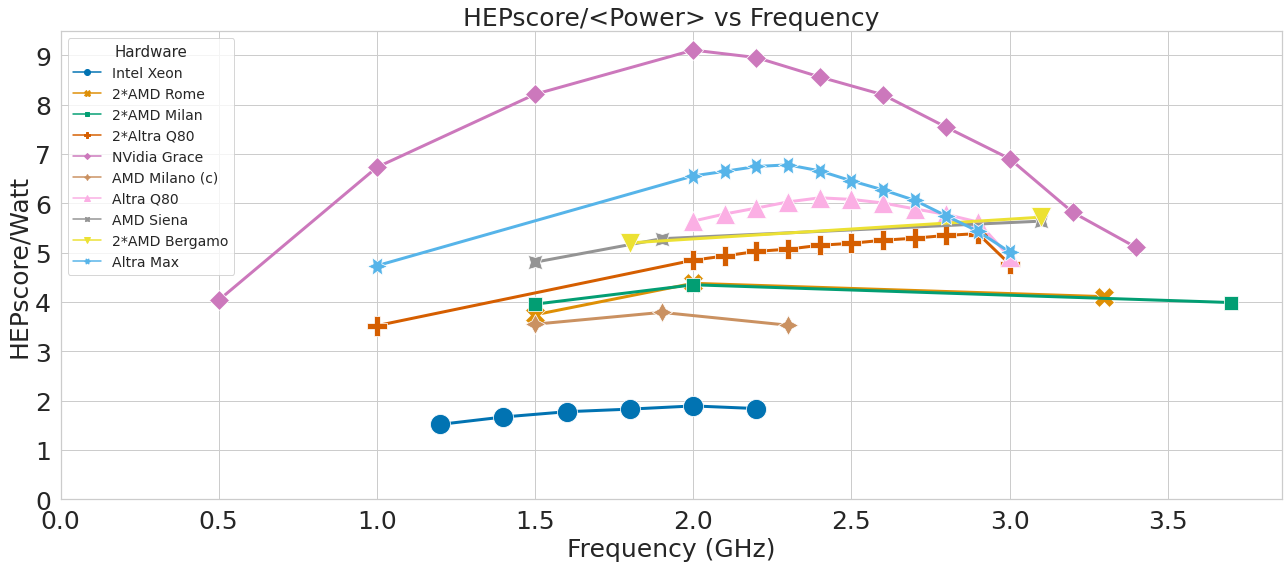}
    \caption{HEPScore/Power vs Frequency. Measurements on different hardwares show that almost all servers have a frequency that maximises the HEPScore/Watt.}
    \label{fig:power-scotgrid}
\end{figure}

Another use case of the enhanced HEP Benchmark Suite is power measurements in conjunction with frequency and HEPScore. Due to the permissions needed for ipmitools \cite{ipmitool}, power measurements cannot be automatically performed on the grid. However, the enhanced suite can be used on bare-metal nodes to conduct such tests and analyses. The measurements presented in \textbf{Figure~\ref{fig:power-scotgrid}} show that nearly all different hardwares have a frequency that maximises the HS23/Watt. These findings can be applied to develop best practices for server configuration. Additionally, this study enables the assessment of machine sustainability across different architectures and hardware setups. This research is ongoing and will continue to provide valuable insights into optimizing performance and energy efficiency. For further details on power efficiency and frequency analysis, see \cite{refId0}. 

\section{Conclusions}
The enhanced HEP Benchmark Suite is versatile across various workload management systems, as demonstrated in ATLAS and LHCb experiments. With plugins enabling measurements of user-defined metrics such as machine load, memory usage, and power consumption, it facilitates detailed performance analysis. The Suite offers valuable insights into the configuration factors that most significantly affect system performance. This includes insights into optimizing server operation, achieving improvements of up to 66\% through configuration adjustments. In addition, the expansion of newer architectures such as ARM \cite{architecture:arm} has arrived, and the system is ready to collect data from it. Last but not the least, HEPScore was used to perform a CPU Consumption Study and test whether it is possible to decrease the power consumption by reducing the CPU frequency. Our measurements highlight that servers can operate at frequencies that maximise the HEPScore/Watt, suggesting better practices for server configuration and enhancing sustainability assessments across diverse hardware setups. The HEPScore with the Suite Plugins is an extremely powerful tool and it can, and should be employed to aid in monitoring experiments compute resources. Its utilisation helps provide a better understanding of the compute resources, detects and enables misconfiguration issues to be rectified, and thus enhances performance as demonstrated by the work described here. Ongoing investigations are exploring additional metrics to better understand performance dynamics, particularly related to CPU frequency and thermal effects.
\bibliographystyle{unsrtnat}
\bibliography{ACAT.bib}  % The name of your .bib file without the .bib extension
\end{document}